\documentclass{svjour2}                    
\smartqed  
\usepackage{graphicx}
\journalname{Journal of Statistical Physics}

\newcommand{\dd}{\mathrm{d}}

\begin{document}

\title{Correlations between record events in sequences of random variables with a linear trend
}


\author{Gregor Wergen \and Jasper Franke \and Joachim Krug}


\institute{G. Wergen \and J. Franke \and J. Krug \at
              Institute for Theoretical Physics, University of Cologne \\
              Tel.: +49 221 7396\\
	      Fax: +49 221 5159\\
              \email{gw@thp.uni-koeln.de, jfranke@thp.uni-koeln.de, krug@thp.uni-koeln.de}           
}

\date{Received: date / Accepted: date}

\maketitle

\begin{abstract}
The statistics of records in sequences of independent, identically distributed random variables is a classic subject of study. One of the earliest results concerns the stochastic independence of record events. Recently, records statistics beyond the case of i.i.d. random variables have received much attention, but the question of independence of record events has not been addressed systematically. In this paper, we study this question in detail for the case of independent, non-identically distributed random variables, specifically, for random variables with a linearly moving mean. We find a rich pattern of positive and negative correlations, and show how their asymptotics is determined by the universality classes of extreme value statistics. 
\keywords{Records \and Extreme Value Statistics \and Correlations}
\end{abstract}

\section{Introduction}
\label{intro}

The interest in studying record events in a general sense is evident as they by definition tend to be rare events of a special nature. As examples of record-breaking events most readers might think about athletic records like a world record in 100 m dash, or long jump \cite{Gembris2002,Gembris2007}. However records also play an important role in other naturally occurring and man-made phenomena.
In particular, climatic records attract much attention and are frequently mentioned in the media in the context of global warming. The statistics of record temperatures, precipitation amounts and floods have been adressed extensively in recent work \cite{Bassett1992,Benestad2003,Redner2006,Meehl2009,Wergen2010,Newman2010,Anderson2010,DeHaan2006}. Also in evolutionary biology record events have a certain relevance as was pointed out in \cite{Sibani1998,Krug2005,Park2008}, and they play an important role in the dynamics of disordered systems like spin glasses and dirty superconductors \cite{Oliveira2005,Sibani2006}. Last but not least records can be studied in the financial sciences, e.g. when considering record-breaking values or changes in stock prices \cite{Wergen2011}.
In all these examples not only the rate of record events is of interest but also their correlations: Given that a record flood just occurred, can one relax because it will take a while before the next record flood is to be expected? Based on our previous work in \cite{Franke2010}, this question will be addressed in the present paper. 
 
Most previous studies of record events and record values \cite{DeHaan2006,Arnold1998,Nevzorov2001,Glick1978} have been restricted to the case where the 
underlying random variables (RV's) $\{X_{l}\}$ are independently drawn from a common distribution $f(x)$ (i.i.d. RV's). In this case, record events can be shown to be stochastically independent, i.e. the knowledge about previous record events does not indicate whether a future record event is more or less likely. However in many applications, the underlying random variables are not i.i.d. For example, in recent work Majumdar and Ziff computed the statistics of records in a symmetric random walk \cite{Majumdar2008} (see also \cite{Wergen2011,LeDoussal2009,Sabhapandit2010}). Here, record events are obviously not independent: In a random walk the probability for a second record to occur immediately after after a previous record is simply $1/2$, whereas the unconditional probability for a record in the 
$N$'th step is $\approx \frac{1}{\sqrt{\pi N}}$ \cite{Majumdar2008}.

In this article we consider correlations between records in sequences of RV's that are independent but not identically distributed. An important example is the statistics of hot or cold records in weather data. In \cite{Wergen2010} the occurrence of record-breaking temperatures in European and American weather recordings was analyzed by modeling daily temperature measurements as uncorrelated RV's with a linear drift. It was shown that the increase of hot records and the decrease of cold records due to global warming can be described and quantified by this model. We will refer to this particular example of non-i.i.d. RV's,  which was originally introduced in the 1980's in the context of athletic records \cite{Ballerini1985,Ballerini1987}, as the Linear Drift Model (LDM). 

The precise definition of the LDM is as follows: For a positive constant $c$, each random variable $Y_l$ has an i.i.d. part $X_{l}$ with density $f(x)$ and an additive linear drift such that
\begin{equation}
\label{LDM}
Y_{l}=X_{l}+cl.
\end{equation}
In other words, while the shape of the density $f_{l}(y)$ of the $l^{th}$ RV is invariant, its overall position is shifted by a constant amount $c$ each time a new RV is drawn.
As was already noted in \cite{Franke2010}, the record events under this model are not in general independent (see also
\cite{Krug2007} for the related case of RV's with changing variance). Below we shall see that the sign and magnitude
of correlations depend in a highly non-trivial way on the underlying probability density $f(x)$. 
To quantify the correlations, consider the joint probability $p_{N, N-1}$ of a record occurring in the $(N-1)^{th}$
as well as in the $N^{th}$ step,
\begin{eqnarray}
 p_{N, N-1}\equiv \mathrm{Prob}[Y_{N} \;\; \mathrm{record} \;\; \mathrm{and} \;\; Y_{N-1} \;\; \mathrm{record}].
\end{eqnarray}
 To normalize this joint probability, we divide by the probabilities  $p_{N}$ and $p_{N-1}$ of records in the 
$N^{th}$ and ${N-1}^{th}$ steps, obtaining the key quantity under consideration throughout this paper, 
\begin{equation}\label{lnn}
  l_{N, N-1}(c)\equiv \frac{p_{N, N-1}}{p_{N}p_{N-1}}.
\end{equation}  
\begin{figure}
  \includegraphics[scale=0.9]{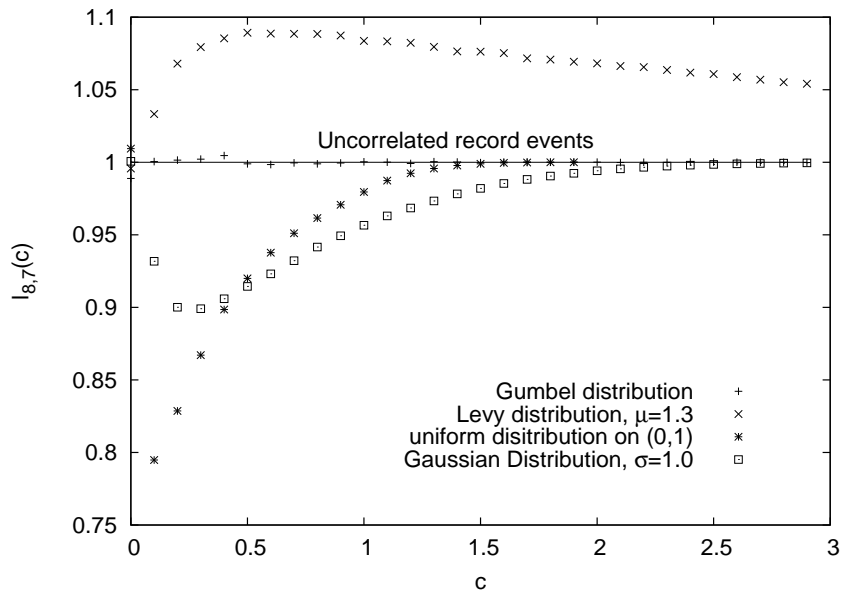}
  \caption{\label{rec_dep} Correlations between record events in the
    $7^{th}$ and $8^{th}$ RV for different distributions of the
    underlying i.i.d. part. For the Gumbel case, the well known
    stochastic independence of record events is found, while
    simulations for the other distributions show that generally the records are correlated.
    Note in particular the curve for the symmetric L\'evy-stable distribution with index $\mu = 1.3$,
    which indicates an attraction between record events.}
\end{figure}
The simulations shown in Fig. \ref{rec_dep} illustrate two known results, the
fact that record events for i.i.d. RVs are stochastically independent \cite{Arnold1998,Nevzorov2001,Glick1978}
(all curves meet at unity for $c\equiv 0$), and that for the LDM with Gumbel distributed
i.i.d. part, record events are stochastically independent for all values
of $c$ \cite{Franke2010,Nevzorov2001,Ballerini1985,Ballerini1987}. 
However the curves for other distributions show that stochastic independence does not hold
in general. For some choices of the underlying
probability density $f(x)$ record events tend to repel each other
($l_{N, N-1}(c)<1$), wheras in other cases the correlations are
positive, i.e. record events attract each other ($l_{N, N-1}(c)>1$). In particular this last,
somewhat counterintuitive numerical finding triggered our interest in this
problem.

In the next section we will derive a general expression for the correlations between record events as quantified by $l_{N, N-1}(c)$. 
For this we employ methods similar to those used in \cite{Franke2010}. 
In Section \ref{examples}, we provide some explicit examples illustrating the rich and counterintuitive behavior of the correlations. 
In particular, we discuss the behavior of correlations with respect to the three classes of extreme value statistics \cite{DeHaan2006,Gumbel1954,Galambos1987,Sornette}. We find that while in the Weibull class of distributions with finite support the correlations are generally negative, positive correlations mainly arise when the RV's are in the Frech\'et class of distributions with power law tails. In the intermediate Gumbel class of distributions with exponential-like tails correlations can be both positive and negative. Finally in Section \ref{conclusions}, we summarize our
results and discuss possible applications and open questions. 

\section{General theory}
\label{general}

Consider a sequence $\{ X_l \}$ of independent but not necessarily identically distributed RV's, with
$f_l(x)$ denoting the probability density of $X_l$.  
The probability that the maximum of the first $N$ RV's is smaller than a 
given value $x$ then factorizes into the probabilities of the RV's
individually being less than $x$,
\begin{equation}\label{factor}
  \mathrm{Prob}[\max\{X_1,...,X_N\} \leq
  x]=\prod_{l=1}^{N}\int^x\dd x' f_{l}(x')=\prod_{l=1}^{N}F_l(x), 
\end{equation}   
 where $F_l(x)=\int^x\dd x' f_{l}(x')$ denotes the cumulative distribution function 
 for the density $f_l(x)$, and the omitted lower (upper) integral
 boundary is understood to be the lower (upper) bound of the support
 of $f_l$ respectively. Throughout we assume distribution functions to be continuous. 

Using Eq. \ref{factor} the probability $p_{N}$ that the $N^{th}$ RV is a record
can be written as
\begin{equation}\label{single}
  p_{N}=\int \dd x f_{N}(x)\prod_{l=1}^{N-1}F_l(x),
\end{equation}
and the joint probability
that both $X_{N-1}$ and $X_{N}$ are records is
\begin{equation}\label{joint}
  p_{N,N-1}=\int \dd x_{N}f_{N}(x_{N})\int^{x_{N}}\dd
  x_{N-1}f_{N-1}(x_{N-1})\prod_{l=1}^{N-2}F_{l}(x_{N-1}). 
\end{equation}
For the LDM we have $f_l(y) = f(y-cl)$ and $F_l(y) = 
F(y-cl)$, where $f(x)$ and $F(x)$ are the density and distribution function of the 
i.i.d. part $X_l$ in (\ref{LDM}). It then follows that \cite{Franke2010}
\begin{equation}
\label{pN_LDM}
p_N = \int \dd y \; f(y-cN) \prod_{l=1}^{N-1} F(y-cl) = \int \dd x \; f(x) \prod_{l=1}^{N-1} F(x+cl) 
\end{equation}
and similarly
\begin{equation}
\label{joint_LDM}
p_{N,N-1} = \int \dd y \; f(y) \int^{y+c} \dd x \; f(x) \prod_{l=1}^{N-2} F(x+cl).
\end{equation}

\subsection{Explicit stochastic independence for two cases}
The expressions (\ref{joint}) and (\ref{joint_LDM}) allow
us to verify two known results: Both for the i.i.d. case
($c\equiv 0$) and for the LDM with Gumbel distributed i.i.d.
part $X_l$, record events are independent.

We start with the i.i.d. case. First note that the probability that the
$N^{th}$ RV is a record is the probability that it is the largest among
$N$ equivalent RV's, so by symmetry $p_{N}(c=0)=1/N$. For the i.i.d.
case, the subscripts in (\ref{joint}) can be dropped. Using the
substitution $u=F(x)$ and $\dd u=f(x) \dd x$, one obtains 
\begin{equation}\label{iid_sti}
  p_{N, N-1}(c=0) 
    =  \int_{0}^{1}\dd u\int _{0}^{u}\dd u'
  u'^{N-2} = \frac{1}{N-1}\frac{1}{N} =p_{N}p_{N-1}.
\end{equation}
It is possible to insert an arbitrary spacing between the two
record events and still obtain this factorization to yield $p_{N,
  N-j}=p_{N}p_{N-j}$. Similarly, one can consider more than one record
event. The factorization property implies that in the i.i.d. case,
record events are stochastically independent \cite{Arnold1998,Nevzorov2001,Glick1978}.

Next we consider the LDM with Gumbel distributed i.i.d. part,  
\begin{equation}
\label{Gumbeldist}
f(x)=\exp\left(-e^{-x}-x\right) 
\end{equation} and
$F(x)=\exp\left(-e^{-x}\right)$. Using that in this case
\cite{Gumbel1954}
\begin{equation}\label{gf}
  F(x+a)=\exp\left(-e^{-x-a}\right)=\exp\left(-e^{-x}\right)^{e^{-a}}
=F(x)^{e^{-a}},
 \end{equation}
the substitution $u=F(x)$ used above yields \cite{Franke2010,Ballerini1985} 
\begin{equation}
p_{N}(c)=\int \dd x f(x) \prod_{l=1}^{N-1} F(x)^{e^{-cl}} =\int_{0}^{1} \dd u
\; u^{\sum_{l=1}^{N-1}\alpha^{l}}=\frac{1}{\sum_{l=0}^{N-1}\alpha^{l}}
\end{equation}
with $\alpha = e^{-c}$, and
\begin{eqnarray}\label{gumbel_sti}
  p_{N, N-1}(c) & = & \int_{0}^{1}\dd u \int_{0}^{u^{\alpha}}\dd u'
  u'^{\sum_{l=1}^{N-2}\alpha^l}=\frac{1}{\sum_{l=0}^{N-2}\alpha^{l}}\int_{0}^{1}\dd
  u \; u^{\sum_{l=1}^{N-1}} \nonumber \\
   & = &
   \left(\frac{1}{\sum_{l=0}^{N-2}\alpha^l}\right)\left(\frac{1}{\sum_{l=0}^{N-1}
       \alpha^l}\right) =p_{N-1}p_{N}.  
\end{eqnarray}
Again one can introduce an arbitrary number of spaces and 
intermediate records and still have the corresponding joint
probability factorize. This implies that for the LDM with Gumbel distributed i.i.d.
part, as for arbitrary distributions without drift, record events
are stochastically independent.

\subsection{Expansion for small drift}
\label{Sec:Expansion}
As shown above, for the i.i.d. case $c \equiv 0$, record events are independent. 
On the other hand, for very large drift, almost
every RV is a record, and thus trivially
$l_{N, N-1}(c)\to 1$ for $c \to 1$ (see Fig. \ref{rec_dep}). 
The behavior of $l_{N,N-1}(c)$ in between these two limits is not as easily
characterized. In particular, while for some probability densities
the records seem to repel each other for intermediate $c$ ($l_{N,
  N-1}(c)<1$), for other  probability densities the records seem to
\emph{attract} ($l_{N, N-1}(c)>1$). 

\begin{figure}
  \includegraphics[scale=0.9]{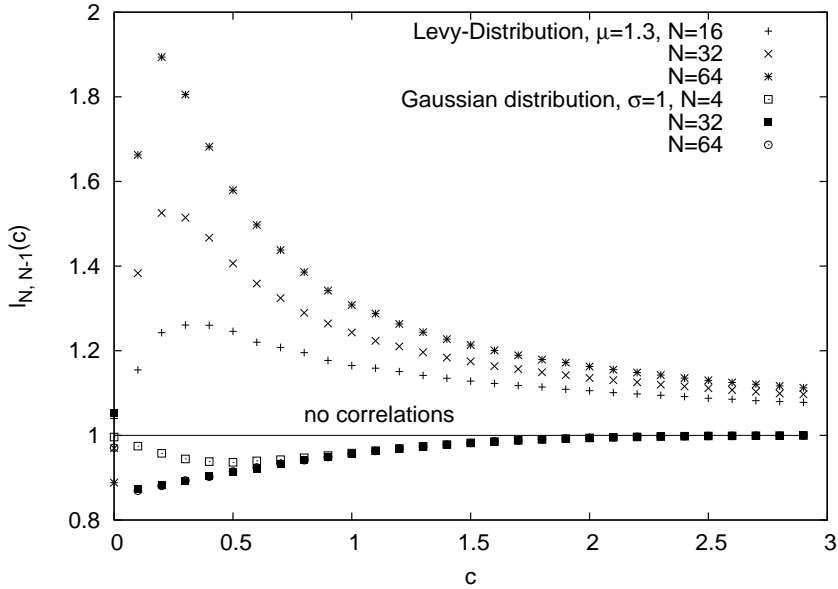}
  \caption{\label{no_fse} Correlations between record events as the
    number of RV's $N$ increases. For the i.i.d. part drawn from a Gaussian
  distribution with variance $\sigma=1$, the correlations are negative and for
  sufficiently large $N$ tend to a limiting form. For the symmetric L\'evy
  distributed i.i.d. part, the correlations are positive and increase
  with $N$. }
\end{figure}

Figure \ref{no_fse} shows examples for both types of behavior, illustrating in addition
how the correlations evolve with increasing $N$ for the cases of L\'evy and Gauss
distributed i.i.d. parts. Clearly, the correlations are not a finite size effect, but their $N$-dependence is markedly different in the two cases. 
For the Gaussian case, the correlations are negative and
seem to approach a form independent of $N$. For the L\'evy case, on
the other hand, the correlations seem to increase with $N$. We will argue later that 
the correlations eventually become $N$-independent also for heavy-tailed distributions,
but this limit is approached much more slowly than in the Gaussian case.

A first step towards understanding the behavior of the correlations
is to determine the sign with which the curves in Figs.\ref{rec_dep} 
and \ref{no_fse} depart from the i.i.d. case $c\equiv 0$. 
To address this question, we compute the record rate
$p_{N}(c)$ and the joint probability 
$p_{N, N-1}(c)$ in the limit of small $c$ and $c/\sigma \ll N^{-1}$, where $\sigma$ is the standard deviation or some other measure of the width of the distribution.
For the record rate this problem was discussed extensively in \cite{Franke2010} (see also \cite{Wergen2010}). Knowing that $p_{N}(c=0)=1/N$, a Taylor expansion for small $c$ yields 
\begin{equation}
p_N\left(c\right) \approx  \frac{1}{N} +
c\frac{N\left(N-1\right)}{2} I(N-2),.
\end{equation}
where we have defined the quantity\footnote{Note that this definition differs slightly from that of the related quantity $I_n$ used in \cite{Franke2010}.
The integral (\ref{I(N)}) also appears in the analysis of the density of near-extreme events \cite{Sabhapandit2007}. }
\begin{equation}
\label{I(N)}
I\left(N\right) =\int\mathrm{d}x f^2\left(x\right) F^N\left(x\right).
\end{equation}
In the same small $c$ regime we can also easily determine the joint probability of occurrence of two consecutive records $p_{N,N-1}\left(c\right)$. To leading order in $c$, the expansion of 
Eq.(\ref{joint_LDM}) yields
\begin{eqnarray}\label{pnm_taylor}
 p_{N,N-1}\left(c\right)  
& \approx & \int \mathrm{d}y f\left(y\right)\int^y \mathrm{d}x
f\left(x\right) F^{N-2}\left(x\right) + c\int \mathrm{d}y f^2\left(y\right)F^{N-2}\left(y\right)
  \nonumber \\  
& & + c\frac{\left(N-1\right)\left(N-2\right)}{2} \int \mathrm{d}y f\left(y\right)\int^y \mathrm{d}x f^2\left(x\right) F^{N-3}\left(x\right).\end{eqnarray}  
The first term can be evaluated by partial integration to
$\frac{1}{N\left(N-1\right)}$,
and another partial integration yields
\begin{eqnarray}\label{pnm}
 p_{N,N-1}\left(c\right) & \approx & \frac{1}{N\left(N-1\right)} -
 c\frac{\left(N-1\right)\left(N-2\right)-2}{2}I\left(N-2\right) \\  
& + & c\frac{\left(N-1\right)\left(N-2\right)}{2}I\left(N-3\right).
\end{eqnarray}
In addition to the zeroth order term $\frac{1}{N\left(N-1\right)}$, we find
a positive and a negative correction term that describe the effect of the
drift. Depending on which one of these two terms is
larger, positive or negative correlations will result.

A similar expansion can be set up for the normalized correlation
$l_{N, N-1}(c)$ defined in Eq.(\ref{lnn}). Writing 
\begin{equation}\label{taylor_end}
l_{N, N-1}(c) =  1+cJ(N) + {\cal{O}}(c^2),
\end{equation}
a straightforward computation yields
\begin{eqnarray}\label{taylor_mid}
J(N) 
 & = &
 N(N-1)I(N-2)+\frac{N(N-1)^2(N-2)}{2}\left[I(N-3)-I(N-2)\right]\nonumber\\  
 & & -\frac{N^2(N-1)}
{2}I(N-2)-\frac{(N-1)^2(N-2)}{2}I(N-3)  
\end{eqnarray}
Again, both positive and negative terms occur in
(\ref{taylor_mid}), the relative magnitude of which 
depends explicitly on the underlying probability density $f(x)$. 
Thus from (\ref{taylor_mid}), it is not at all obvious how
the counterintuitive positive correlations observed numerically arise,
how their occurrence depends on the properties of $f$ and how the
clustering of record events behaves as a function of the number $N$ of
RV's considered. Before turning to the explicit evaluation of  
$J(N)$ in Section \ref{examples}, we provide some heuristics
on the asymptotics of the correlations in the limit of very large
$N$, and address the behavior of correlations between RV's that are more 
than one time step apart.

\subsection{Asymptotics for large $N$} 
\label{Sec:LargeN}
It is clear from Eqs.~(\ref{pN_LDM}) and (\ref{joint_LDM}) that the asymptotic behavior of the record 
rate $p_N$ and the joint probability $p_{N,N-1}$ hinges on the existence of the function
\begin{equation}\label{g_limit}
G_c(x)\equiv\lim_{N\to\infty}\prod_{j=1}^{N}F(x+cj)
\end{equation}
which was rigorously examined by Ballerini and Resnick. In \cite{Ballerini1985}, 
they prove that the limit (\ref{g_limit}) exists and is nonzero
whenever the density $f(x)$ has a finite first moment. For completeness we provide
a heuristic version of their argument. 
Taking the logarithm of Eq. (\ref{g_limit}) one has to consider the convergence of the series
\begin{equation}\nonumber
  \ln\left(G_c(x)\right)=\sum_{j=1}^{\infty}\ln\left(F(x+cj)\right).
\end{equation}
Now $F(x+cj)=\mathrm{Prob}(X\leq x+cj)=1-\mathrm{Prob}(X>x+cj)$ and for any finite
 $x$ and $c>0$ there is a $\tilde{j}$ large enough such that
 $\textrm{Prob}(X>x+c\tilde{j})\ll 1$. Therefore
\begin{equation}
\label{Rest}
  \ln\left(G_c(x)\right)\approx
  s_{\tilde{j}}(x)-\sum_{j=\tilde{j}}^{\infty}\mathrm{Prob}(X>x+cj), 
\end{equation}
where $s_{\tilde{j}}=\sum_{j=1}^{\tilde{j}-1}\ln(F(x+cj)$
is a finite sum and the logarithms in the remaining series were approximated by their Taylor
expansion around unity. The condition for $f(x)$ to have a finite first moment is that
$\mathrm{Prob}(X>x)$ decays faster than $1/x$, which implies convergence of the 
infinite series in (\ref{Rest}) and hence of $G_c(x)$ (in the opposite case
the series diverges to $-\infty$ and $G_c \equiv 0$).
Since $G_c(x)$ is a probability, it
is bounded from above by unity and it follows from (\ref{pN_LDM}) and (\ref{joint_LDM})
that $p_N(c)$ and $p_{N,N-1}(c)$ have finite, non-zero limits for $N \to \infty$. 
The same statement then applies to the ratio (\ref{lnn}). 

Although the focus of this paper is on the case of a positive drift which enhances the occurrence of (upper) records, it is instructive to compare the asymptotic behavior for $c>0$ described above 
to the case $c<0$. For negative drift both the record rate $p_N$ and the joint probability
$p_{N,N-1}$ vanish for $N \to \infty$, and they do so more rapidly than their i.i.d. counterparts. 
The asymptotic behavior of the ratio (\ref{lnn}) is then \textit{a priori} undetermined. We 
consider a simple example where the quantities of interest can be computed explicitly. Let the 
probability density of the i.i.d. variables be given by the negative exponential distribution,
\begin{equation}
\label{negexp}
f(x) = e^{x}, \;\;\; x \leq 0
\end{equation}
and $f(x) = 0$ elsewhere. Then for $c<0$ the integrand in (\ref{pN_LDM}) and (\ref{joint_LDM}) is
of the form 
\begin{equation}
\label{prodexp}
\prod_{l=1}^{K} F(x+cl) = \exp\left[Kx+\frac{c}{2}K(K+1)\right], \;\;\; x \leq 0,
\end{equation} 
and integration yields
\begin{equation}
\label{pNpNN-1}
p_N(c) = \frac{1}{N} \exp\left[\frac{c}{2}N(N-1)\right], \; p_{N,N-1}(c) = \frac{1}{N(N-1)}\exp\left[\frac{c}{2}N(N-1)\right].
\end{equation}
For $c<0$ the record rate is suppressed superexponentially, such that the expected number of records remains
finite for $N \to \infty$ (see also \cite{Franke2010}). However, forming the ratio (\ref{lnn}) we see that
\begin{equation}
\label{lnn_exp}
l_{N,N-1} = \exp\left[-\frac{c}{2}(N-1)(N-2)\right]
\end{equation}
which, in contrast to the case of positive drift, grows without bound for $N \to \infty$. 
We will see in Section \ref{examples} that such a behavior is not restricted to 
this specific example. The 
non-existence of the $N \to \infty$ limit for $c<0$ suggests that the limiting function
\begin{equation}
\label{last}
l^\ast(c) \equiv \lim_{N \to \infty} l_{N,N-1}(c)
\end{equation}
may not be smooth for $c \to 0$, as will be explicitly demonstrated in Section \ref{examples}.

\subsection{Dependence on the distance between record events}
\begin{figure}
  \includegraphics[width=0.48\textwidth]{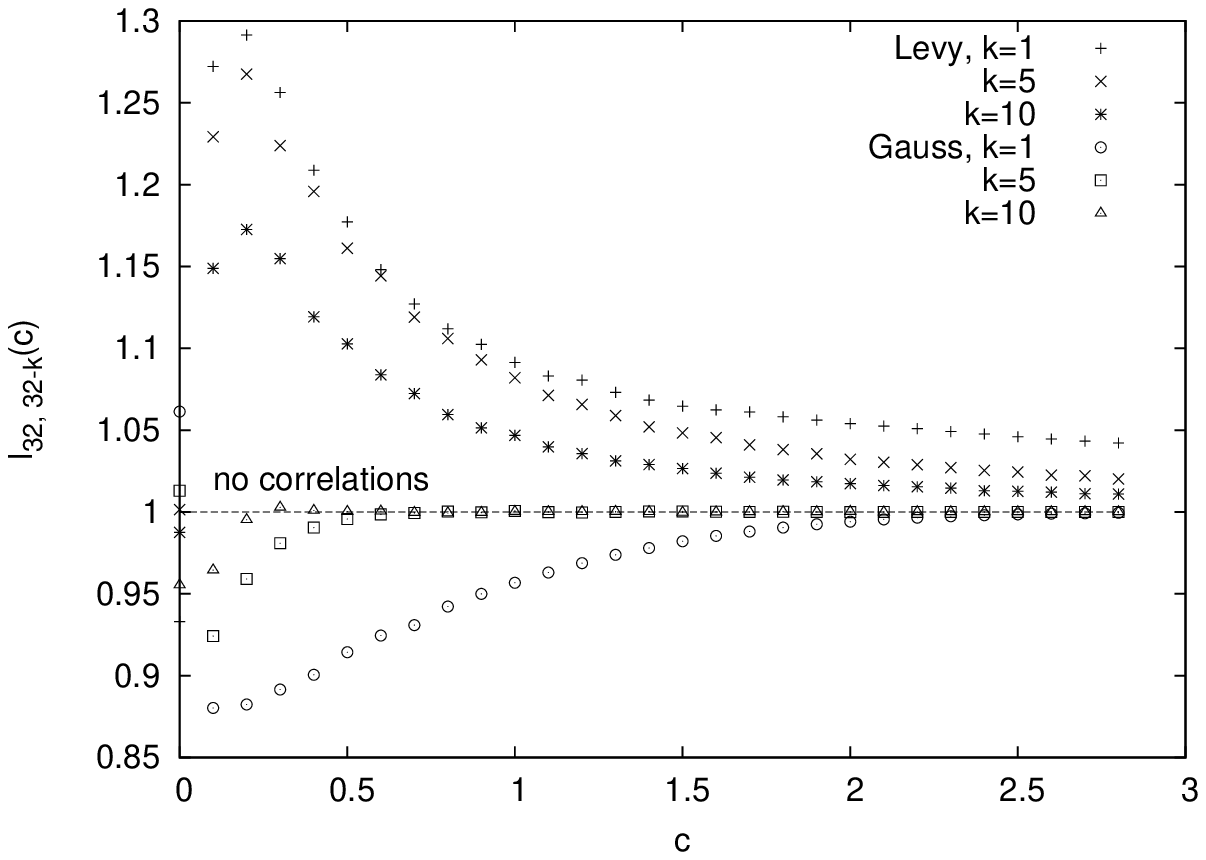}\hfill
  \includegraphics[width=0.48\textwidth]{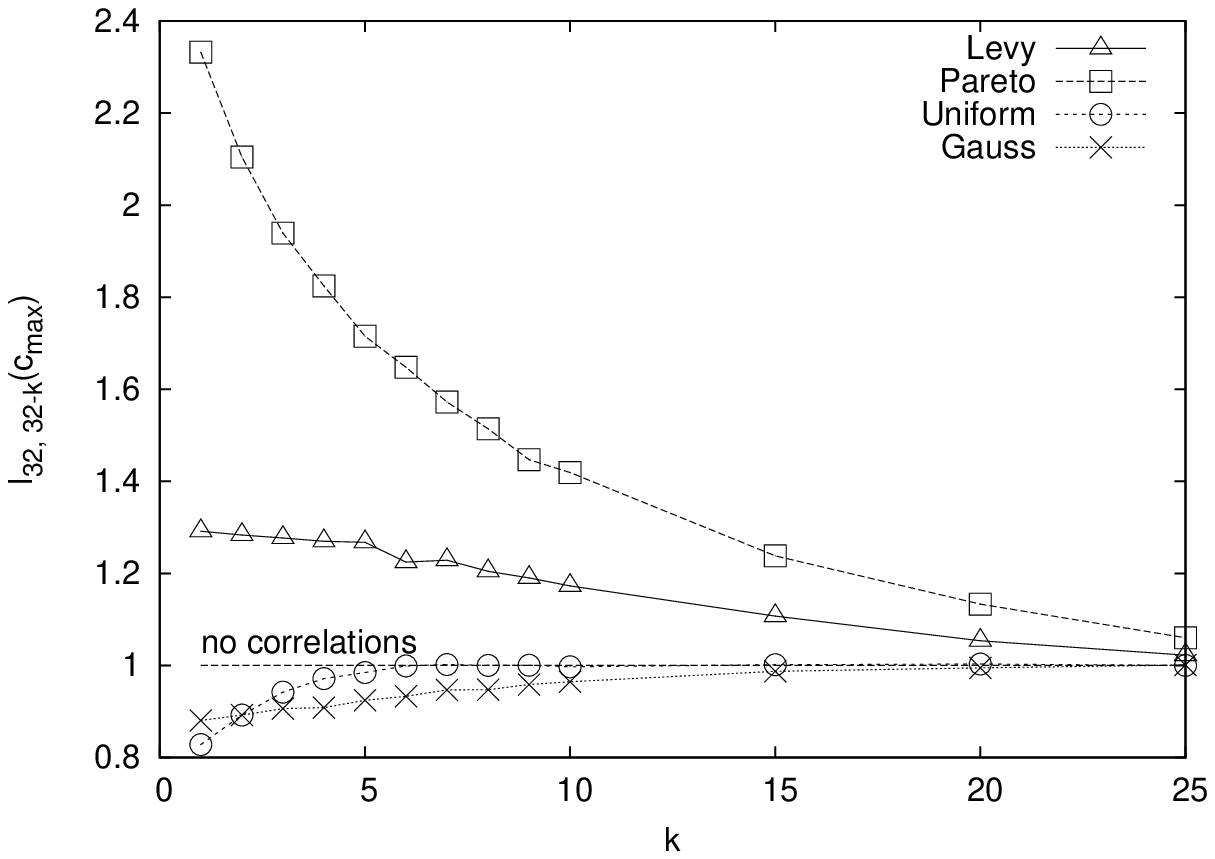}
  \caption{\label{dist_corr} Correlations between record events at
    distance $k$ from time series of length $N=32$ for Gaussian
   ($\sigma=1$), L\'evy-stable ($\mu=1.5$), uniform (on $[0,1]$)
    and Pareto ($\mu=1.5$) distributions. \textbf{Left:} Full $c$
    dependence of $l_{N, N-k}(c)$. \textbf{Right:} At the drift velocity
    $c_{\max}$ where the correlations are maximal ($c_{\max}=0.1$ for
    Gaussian and uniform, $c_{\max}=0.2$ for the L\'evy and
    $c_{\max}=0.4$ for the Pareto case), $l_{N, N-k}(c_{\max})$ is shown
    as function of $k$ to illustrate the decay of correlations.}
\end{figure}

Before embarking on the study of specific distributions, we briefly comment
on the behavior of the correlations as a function of the distance 
between record events. Denoting by $p_{N,N-k}$ the joint probability for observing
a record in the $N-k$th and the $N$th event, it was shown in \cite{Ballerini1985} that
\begin{equation}\nonumber
\lim_{k\to \infty}\lim_{N\to\infty}p_{N, N-k}(c)=\left[\lim_{N \to
    \infty}p_{N}(c)\right]^2 \equiv p(c)^2, 
\end{equation}
which implies that record events become uncorrelated for $k \to \infty$.
This leads to the expectation that in a finite time series,
the correlations between record events at distance $k$ should decay with
$k$, in agreement with the simulations shown in Fig. \ref{dist_corr}. The
correlations are seen to be maximal at $k=1$. In the present paper we therefore focus on
the case of nearest neighbor correlations,
although extending the computations to $k>1$ is
in principle straightforward.

\section{Explicit examples}

\label{examples}

Whereas the statistics of record events in sequences of i.i.d. RV's is completely universal 
\cite{Arnold1998,Nevzorov2001,Glick1978}, in the presence of drift 
one has to distinguish between distributions belonging to the three different 
universality classes of extreme value theory, the 
classes of Weibull, Gumbel and Fr\'echet \cite{DeHaan2006,Galambos1987,Sornette}. 
For each of these classes we will analyze the correlations in the LDM for a few exemplary 
distributions, starting with the Weibull class, and summarize the observed behavior in a unifying scaling picture in Section \ref{Sec:Unified}. The results presented below rely on the study of the record rate $p_N\left(c\right)$ in the LDM presented in \cite{Franke2010}, as well as on the expansions for $p_{N,N-1}\left(c\right)$ [Eq.~(\ref{pnm})] and $l_{N,N-1}\left(c\right)$ 
[Eqs.~(\ref{taylor_end},\ref{taylor_mid})] derived above in Section \ref{Sec:Expansion}.

\subsection{The Weibull class}

\begin{figure}
\includegraphics[width=0.5\textwidth]{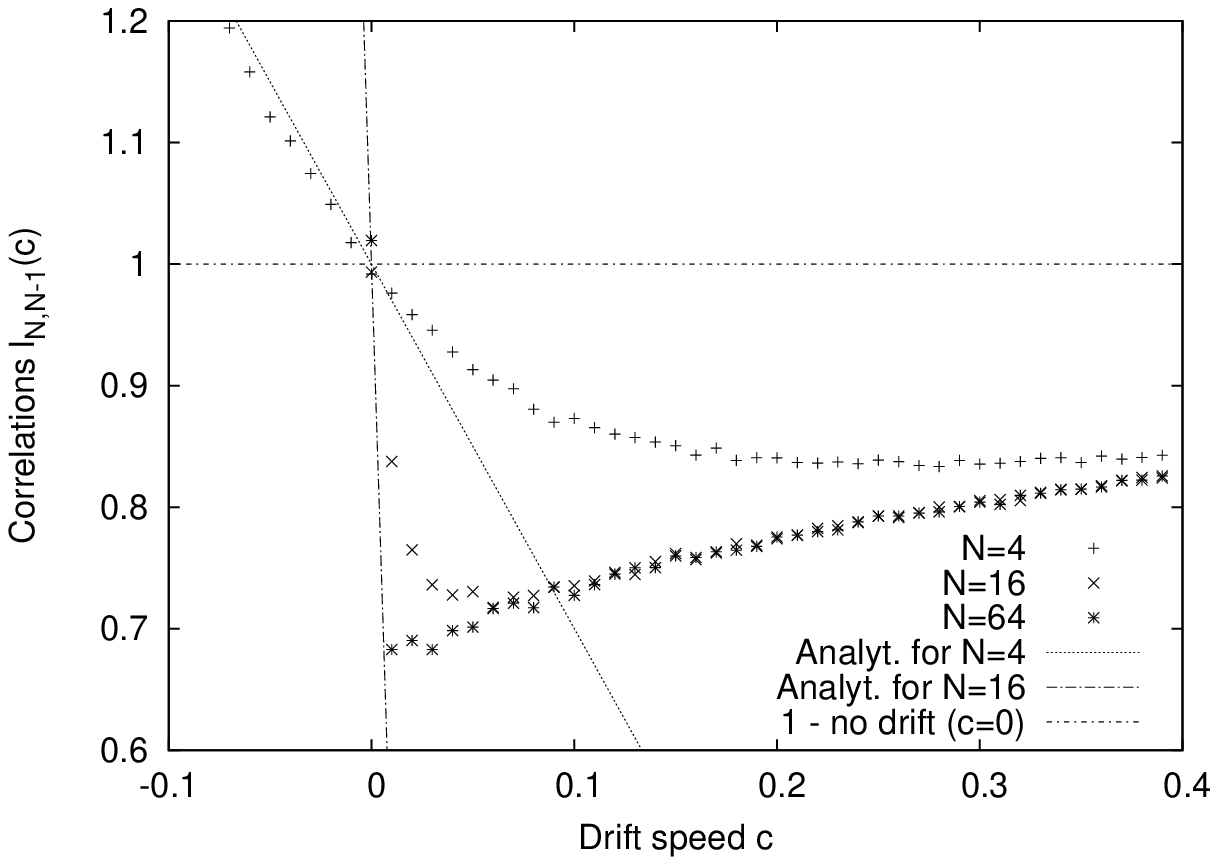}\includegraphics[width=0.5\textwidth]{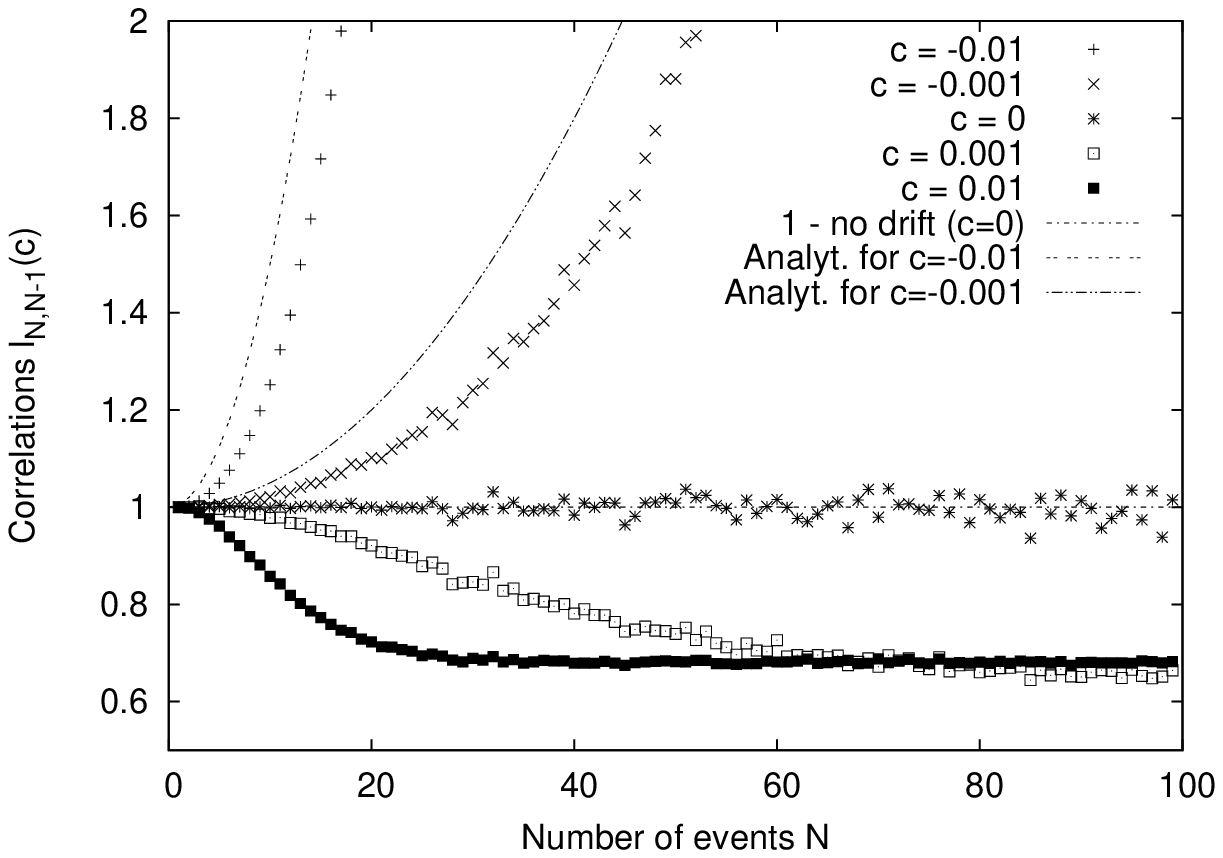}
\caption{\label{Fig:uniform} \textbf{Left:} Numerical simulations of $l_{N, N-1}(c)$ for a uniform distribution of width $a=1$. For each $c$ and $N$ we simulated $10^6$ time series. The figure shows data for $N=4,16$ and $64$. At $c=0$ we find a steep descent of $l_{N, N-1}(c)$. For $N=4$ and $N=16$ we show the analytic prediction  (\ref{prediction_uniform}) (steep lines). 
With increasing $c$, $l_{N, N-1}(c)$ appears to become completely independent of $N$. \textbf{Right:} Simulation results for $l_{N,N-1}$ as a function of $N$ for $c=-0.01,-0.001,0,0.001$ and $0.01$. We simulated $10^7$ series of $100$ RV's in each case. For $c<0$ we find a very steep increase of $l_{N,N-1}$ that is in good agreement with the analytical prediction
(\ref{prediction_uniform}). For $c>0$, $l_{N,N-1}$ quickly saturates at a constant value and our analytical results are not very useful.}
\end{figure}

The Weibull class is the class of distributions with bounded support. As a very simple member of the Weibull class we consider a uniform distribution on the interval $[-a,a]$.
We compute $l_{N,N-1}\left(c\right)$ in the small $c$ regime by making use of 
(\ref{taylor_end},\ref{taylor_mid}) and find for $N\gg1$
\begin{eqnarray}\label{prediction_uniform}
l_{N,N-1}\left(c\right) \approx 1 - \frac{c}{4a}N^2.
\end{eqnarray}
The correlations are negative and their magnitude increases rapidly with $N$
(but note that the leading order approximation is valid only for $cN^2/2a \ll 1$). 
We performed numerical simulations to check this result with $a=1$ and $-0.1<c<0.4$. Figure \ref{Fig:uniform} shows how the descent of $l_{N,N-1}\left(c\right)$ for $c>0$ steepens
with increasing $N$.  
In the figure we also plot our prediction (\ref{prediction_uniform}) for small values of $N$, however the steepness of the descent at zero does not allow us to estimate the quality of our approximation reliably. The figure shows that for large enough $N$ and not too small $c$, $l_{N,N-1}\left(c\right)$ eventually becomes independent of $N$, as expected from the general arguments given in Section 
\ref{Sec:LargeN}. 
In this case the approach to the $N \to \infty$ limit is particularly rapid, because distributions that are more than $2a/c$ steps apart do not overlap. The sharp initial drop of $l_{N,N-1}(c)$ indicates
that the limiting distribution $l^\ast(c)$ is discontinuous at $c=0$, which is consistent with the
non-existence of a limiting function\footnote{With regard to its upper tail, the negative 
exponential distribution (\ref{negexp}) is equivalent to the uniform distribution, and therefore
the argument of Section \ref{Sec:LargeN} applies here as well.} for $c<0$.   

As a more general subset of the Weibull class we next consider the class of 
distributions defined by 
\begin{equation}
\label{Kuramaswamy}
\label{fxi}
f_{\xi}\left(x\right) = \xi\left(1-x\right)^{\xi-1} 
\end{equation}
with $\xi>0$ and $0<x\leq 1$. For $\xi=1$ this produces a uniform distribution similar to the one used above. In \cite{Franke2010} we found that for $\xi>\frac{1}{2}$ the integral in (\ref{I(N)}) is given by
\begin{eqnarray}
I(N)=\xi\frac{\Gamma\left(2-\frac{1}{\xi}\right)\Gamma(N+1)}{\Gamma\left(N+3-\frac{1}{\xi}\right)}.
\end{eqnarray}
Using this and assuming $N\gg 1$ we find the following expression for $l_{N,N-1}\left(c\right)$:
\begin{eqnarray}
 l_{N,N-1}\left(c\right) \approx 1 - \frac{c N^3}{2} \frac{\Gamma\left(2-\frac{1}{\xi}\right)\Gamma\left(N-1\right)}{\Gamma\left(N+1-\frac{1}{\xi}\right)}
\end{eqnarray}
and making use of the Stirling approximation we finally obtain
\begin{eqnarray}
\label{Weibull_gen}
l_{N,N-1}\left(c\right) \approx 1 - \frac{c}{2} \Gamma\left(2-\frac{1}{\xi}\right) N^{1+\frac{1}{\xi}}.
\end{eqnarray}
Similar to the uniform case, the correlations between neighboring record events are always negative 
and their magnitude increases rapidly with increasing $N$, suggesting a discontinuity of the limiting
function $l^\ast(c)$ at $c=0$.

\subsection{The Gumbel class}

\begin{figure}
\includegraphics[width=0.9\textwidth]{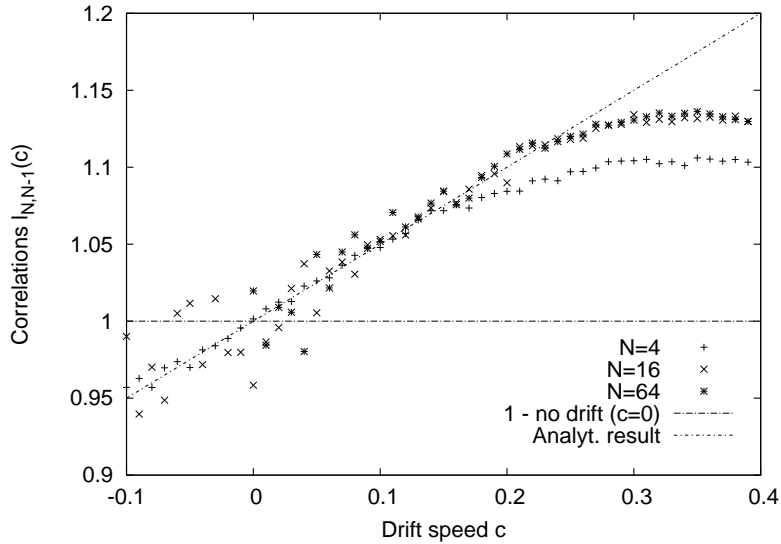}
\caption{\label{Fig:exp} Numerical Simulations of $l_{N, N-1}(c)$ for an exponential distributions with unit mean. For each $c$ and $N$ we simulated $10^6$ time series. The figure shows results for 
$N=4,16$ and $64$ along with our $N$-independent analytical prediction (\ref{lexp}).}
\end{figure}

As a first example in the Gumbel class we consider the exponential distribution with mean 
$a$, 
$f\left(x\right) = a^{-1} e^{-\frac{x}{a}}$, $x \geq 0$. 
In \cite{Franke2010} it was found that in this case the effect of the linear drift on the record rate is independent of $N$ to leading order in $c$. Using this result it is straightforward to obtain
\begin{eqnarray}
\label{lexp}
 l_{N,N-1}\left(c\right) \approx 1 + \frac{c}{2a}.
\end{eqnarray}
We compared this result to numerical simulations in Fig. \ref{Fig:exp} and found a very good agreement. Apparently $l_{N,N-1}\left(c\right)$ assumes a value independent of $N$ already for very small $N$. 

\begin{figure}
\includegraphics[width=0.5\textwidth]{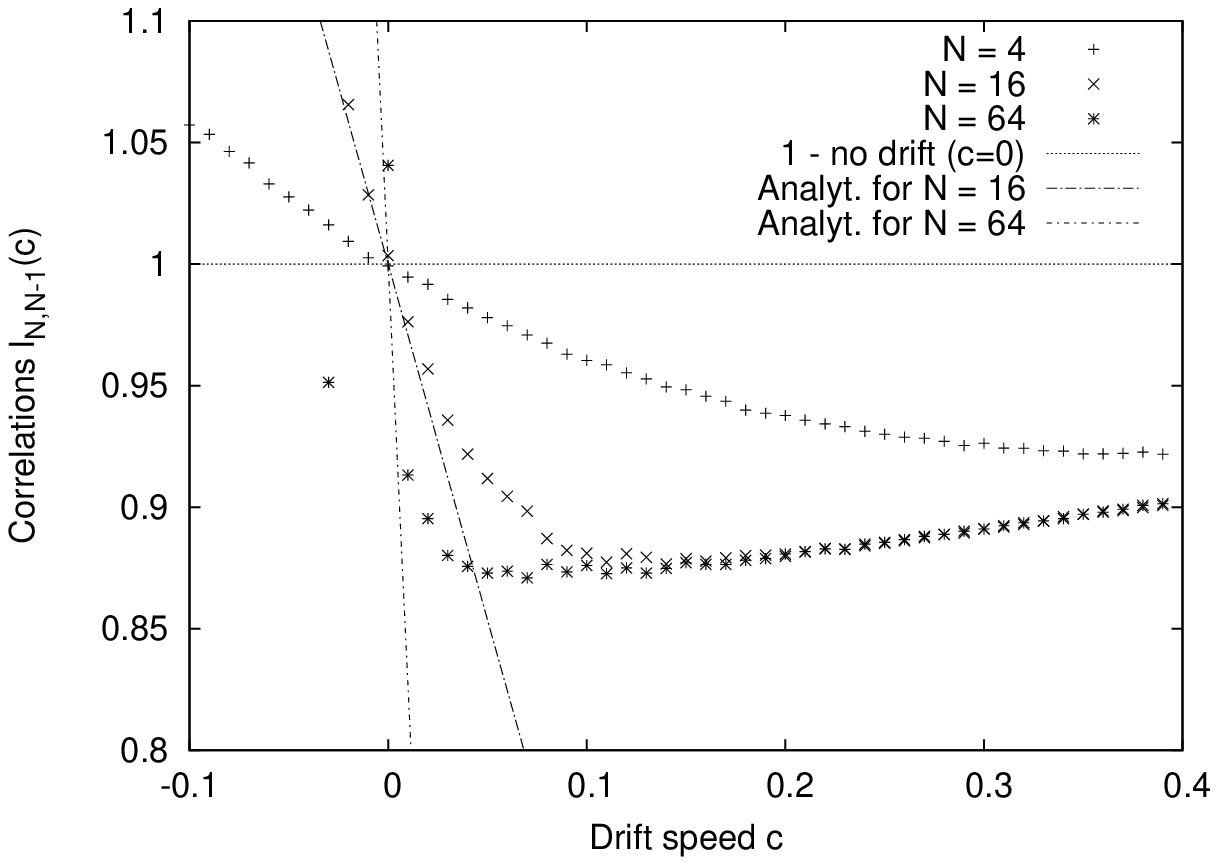}\includegraphics[width=0.5\textwidth]{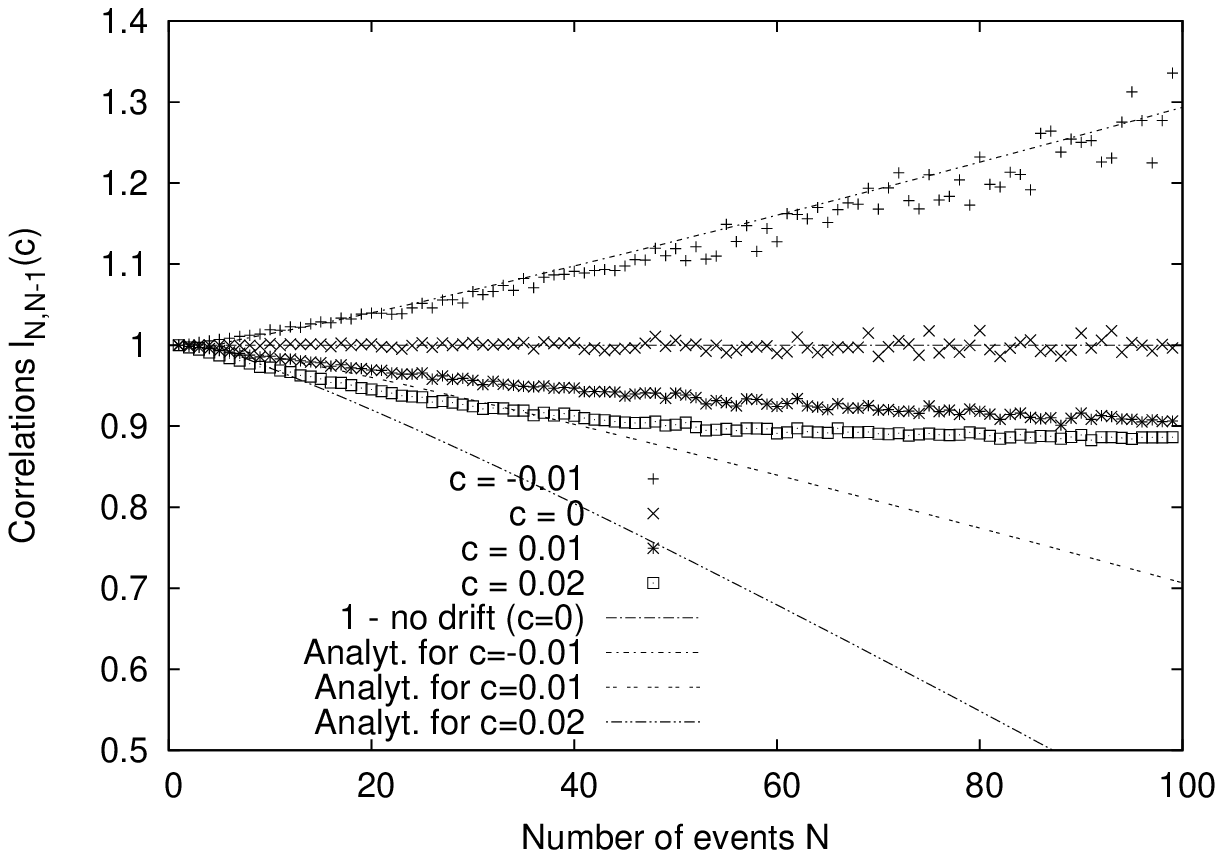}
\caption{\label{Fig:gaussian} Numerical simulations of $l_{N, N-1}(c)$ for a Gaussian distribution of width $\sigma = 1$. \textbf{Left figure} shows $l_{N, N-1}(c)$ for $N=4,16$ and $64$. For each $c$ and $N$ we simulated $10^7$ time series. The figure illustrates the steep descent of the correlator at $c=0$ when $c$ is small. \textbf{Right figure} shows $l_{N, N-1}(c)$ for different, fixed values of $c$ together with our analytical results. Here, we analyzed $10^8$ series of RV's for each drift rate. Again, we find agreement between the simulations and our analytical computations for small $N$ and $c$. We manually fitted curves $\propto N\sqrt{\textrm{ln}\left(N^2/8\pi\right)}$. Interestingly for $c<0$ the agreement is a lot better and $l_{N, N-1}(c)$ does not saturate at a constant value.}
\end{figure}

Another important member of the Gumbel class is the normal distribution with mean zero and standard
deviation $\sigma$. Here, the computation is usually a bit more complicated; however, as in the previous examples, most of the work was already done in \cite{Franke2010}. There we found that for $N\gg1$ and $cN\ll\sigma$
\begin{eqnarray}
 I\left(N\right) \approx \frac{c}{N^2\sigma}\frac{4\sqrt{\pi}}{\textrm{e}^2}\sqrt{\textrm{ln}\left(\frac{N^2}{8\pi}\right)}.
\end{eqnarray}
Using (\ref{taylor_mid})
this result allows us to estimate the effect of the drift on the correlations. We find
\begin{eqnarray}\label{prediction_gaussian}
\left(l_{N,N-1}\left(c\right) - 1\right) \propto - N\frac{c}{\sigma}\sqrt{\textrm{ln}\left(\frac{N^2}{8\pi}\right)}.
\end{eqnarray}
Unlike the case of the exponential distribution, in the Gaussian case the correlations are negative and depend strongly on $N$. The behavior is 
similar to that found in the Weibull case, and matches the result (\ref{Weibull_gen}) (up to logarithimic factors) for $\xi \to \infty$.  
Numerical results for the Gaussian distribution are shown in Fig. \ref{Fig:gaussian}. Similar to the uniform distribution, $l_{N,N-1}\left(c\right)$ shows a steep descent at $c=0$ which becomes steeper for increasing $N$.
However, the descent appears to be smoother than in the uniform case. In agreement with the considerations of Section \ref{Sec:LargeN},
for $c>0$ and large $N$ the correlations become independent of $N$. In contrast, for $c<0$ the correlations increase monotonically with $N$ and 
do not appear to saturate. In the case of the Gaussian distribution our analytical prediction (\ref{prediction_gaussian}) is only valid in a small regime for small $c>0$ and 
relatively small $N$, but for $c<0$ the approximation is significantly better.

To understand the marked difference between the exponential and Gaussian distributions, we analyze the general class of Gumbel-type distributions of the form 
\begin{equation}
\label{fbeta}
f\left(x\right) = A_\beta\textrm{e}^{-|x|^\beta}, 
\end{equation}
with normalization $A_\beta$ and $\beta>0$. In this case it was found in \cite{Franke2010} that 
\begin{equation}
\label{IGumbel}
I\left(N\right) \propto N^{-2}\textrm{ln}\left(N\right)^{1-\beta^{-1}}.
\end{equation} 
With some further computations this leads us to the following behavior of the leading order correction coefficient $J\left(N\right)$:
\begin{eqnarray}
\label{JGumbel}
 J\left(N\right) \propto - D_1 \left(1-\frac{1}{\beta}\right) N \; \textrm{ln}\left(N\right)^{1-\frac{1}{\beta}} + D_2 \; \textrm{ln}\left(N\right)^{1-\frac{1}{\beta}},
\end{eqnarray}
where $D_1$ and $D_2$ are positive constants not depending on $N$. The exact values of $D_1$ and $D_2$ are very difficult to compute and we will not consider them in this article. 
Nevertheless, this expression nicely reproduces our results for the exponential and the Gaussian distribution and shows how both positive and negative correlations may emerge in the Gumbel class. 
For $\beta=1$ the leading first term vanishes and $J(N)$ reduces to a positive constant. For all $\beta\neq1$ the first term dominates and the correlations grow with $N$, with a sign
determined by $1-\frac{1}{\beta}$. For values $\beta>1$ of distributions that decay faster than the standard exponential we find negative correlations between the records, in agreement with the Gaussian example,
while for stretched exponential distributions ($\beta<1$) positive, $N$-dependent correlations result (see Fig.~\ref{Fig:comp} for an example). 

In hindsight, the special role of the exponential distribution should not come as a surprise, since the Gumbel distribution (\ref{Gumbeldist}) (for which correlations are completely absent) has an exponential tail. 
In the sense of extreme value theory, the exponential distribution is close to the Gumbel, and the corresponding records in the LDM are therefore almost uncorrelated (up to small residual correlations which are 
independent of $N$). 

\subsection{The Fr\'echet class}

\begin{figure}
\includegraphics[width=0.9\textwidth]{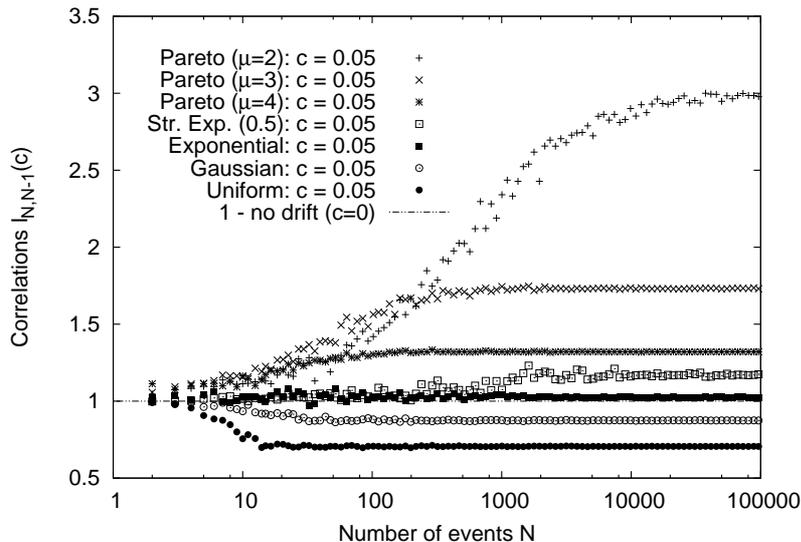}
\caption{\label{Fig:comp} Numerical simulations of $l_{N, N-1}(c)$ with $c=0.05$ for six different distributions. In each case we averaged over $10^6$ time series of length $N=10^5$. The results for $l_{N, N-1}(c)$ were binned logarithmically in order to improve the averaging
for large $N$. For the uniform and the Gaussian distribution $l_{N, N-1}(c)$ is clearly negative for $c>0$ and it decreases with growing $N$. At some $N = N^\ast$ the correlations become independent of $N$. For the special case of the exponential distribution we find a constant value of $l_{N, N-1}(c)$ which is only slightly larger than unity. 
Both the stretched exponential (with $\beta=1/2$) and the Pareto distributions (with $\mu=2,3,4$) show positive correlations. 
Note the slow convergence for the Pareto distribution with $\mu = 2$.} 
\end{figure}

As a subset of the distributions in the Fr\'echet class of extreme value statistics we consider the well known Pareto distribution
\begin{equation}
\label{Pareto}
f\left(x\right) = \mu x^{-\mu-1} 
\end{equation}
with $\mu>1$ and $x\geq1$. 
In \cite{Franke2010} it was found that in this case
\begin{eqnarray}
I(N)=\mu\frac{\Gamma\left(2+\frac{1}{\mu}\right)\Gamma(N+1)}{\Gamma\left(N+3+\frac{1}{\mu}\right)}.
\end{eqnarray}
This allows us to compute $l_{N,N-1}\left(c\right)$ for the Pareto distribution. We find
\begin{eqnarray}
 J(N) =  \frac{\mu }{2} \frac{\Gamma\left(2+\frac{1}{\mu}\right)\Gamma\left(N-1\right)}{\Gamma\left(N+\frac{1}{\mu}+1\right)} \frac{N\left(N-1\right)\left(N^2+\mu N + \mu\right)}{\mu N + \mu + 1}
\end{eqnarray}
\begin{figure}
\includegraphics[width=0.5\textwidth]{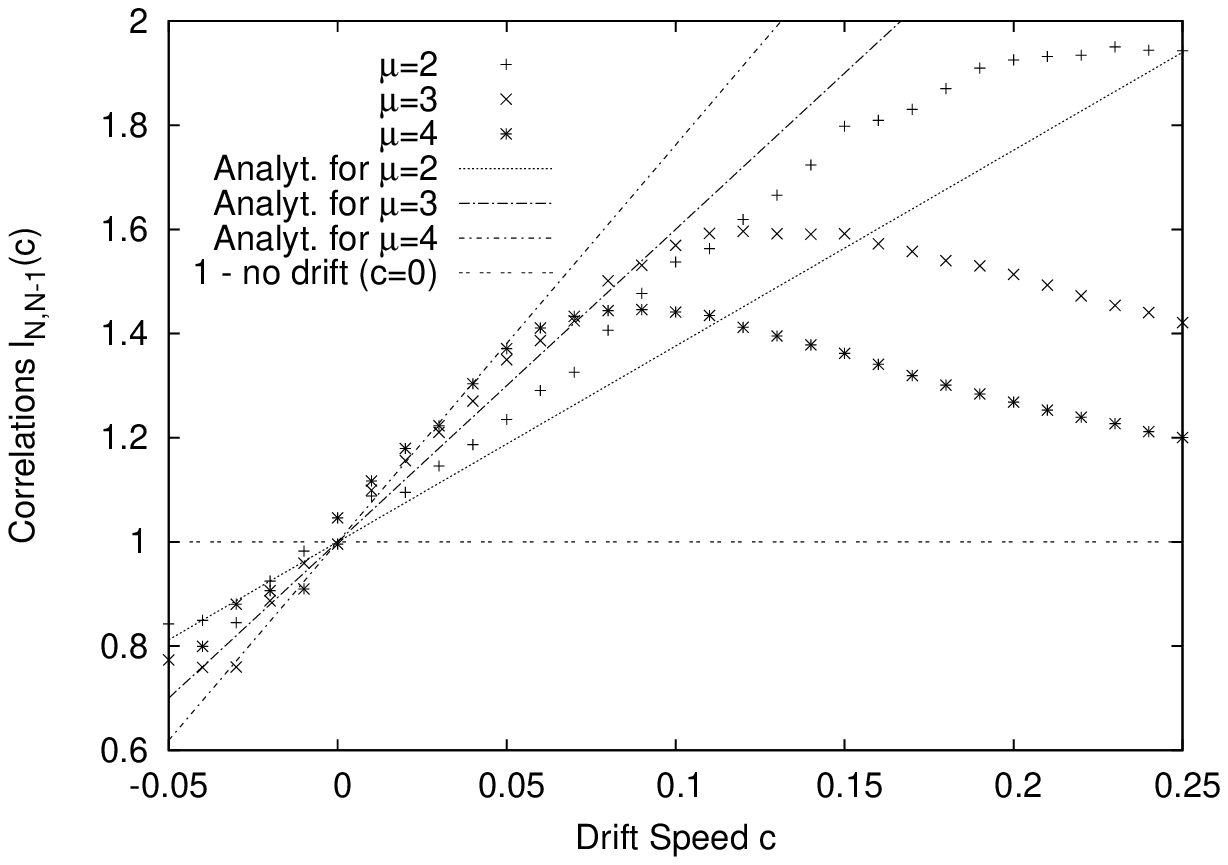}\includegraphics[width=0.5\textwidth]{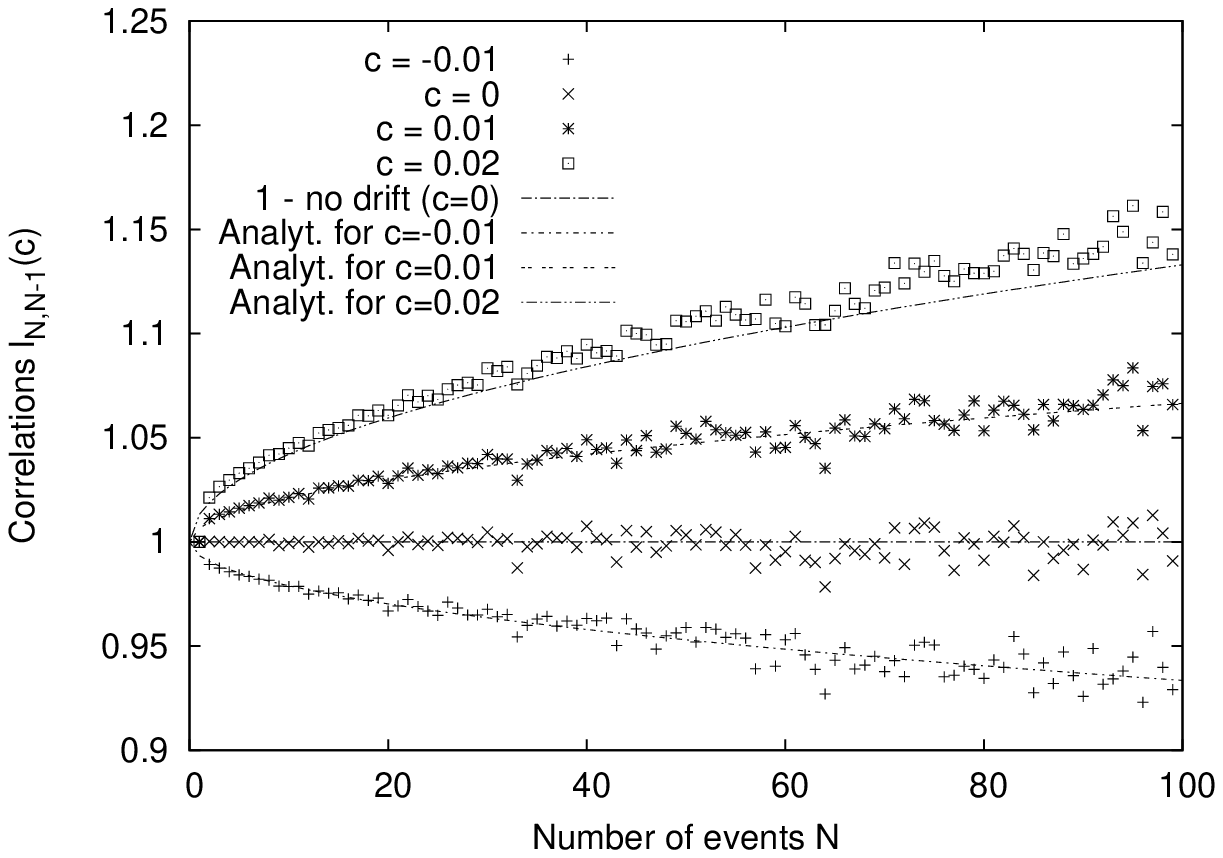}
\caption{\label{Fig:pareto} \textbf{Left:} Numerical simulations of $l_{N, N-1}(c)$ for Pareto distributions. For each $c$ and $\mu$ we simulated $10^6$ time series of length $N=16$. The figure shows simulations for $\mu=2,3$ and $4$, along with our analytical predictions for small $c$. At $c=0$ we find a steep ascent of $l_{N, N-1}(c)$ which gets steeper with increasing $\mu$. Our analytical result from (\ref{prediction_pareto}) is in good agreement with the simulations for small enough $c$. 
\textbf{Right:} Simulation results for a Pareto distribution with $\mu=2$ and different values of $c$. We analyzed $10^8$ series of $N=100$ RV's for each $c$. Our analytical work predicts a square-root 
behavior of $l_{N, N-1}(c)-1$, which is in good agreement with the simulations. }
\end{figure}
which leads to the expression
\begin{eqnarray}\label{prediction_pareto}
 l_{N,N-1}\left(c\right) \approx 1+ \frac{c}{2}\Gamma\left(2+\frac{1}{\mu}\right)N^{1-\mu^{-1}}
\end{eqnarray}
for large $N$.
For positive $c$, we thus expect positive correlations between neighbouring record events for all distributions of Pareto form.
These correlations are increasing with $N$ slower than linearly. We expect that, asymptotically, this behavior is universal for all distributions within the Fr\'echet class\footnote{In \cite{Franke2010} we
presented data for a L\'evy stable distribution with $\mu = 1.8$ 
which show negative correlations between record
events. We have checked that this behavior is not asymptotic, and that the correlations become
positive for larger values of $N$.}. We compare Eq.(\ref{prediction_pareto}) to numerical simulations in Fig. \ref{Fig:pareto}, finding good agreement for small $c$, both for the $c$- and the $N$-dependence. The agreement improves when the distribution becomes broader for smaller $\mu$. The strength of correlations for strongly heavy-tailed distributions of the Pareto class is remarkable. For example, for a Pareto distribution with coefficient $\mu=2$ we found correlations $l_{N,N-1} \approx 3$ for large $N$ (see Fig.~\ref{Fig:comp}).

The correlations displayed in the right panel of Fig.\ref{Fig:pareto} increase (for $c > 0$) or 
decrease (for $c < 0$) with $N$ without showing any sign of saturation, although 
we know from the general considerations of Section \ref{Sec:LargeN} that $l_{N,N-1}$ must approach
an $N$-independent limit for $c > 0$ and $\mu > 1$. 
As we will see in the next subsection, this reflects the
fact that the time scale $N^\ast$ at which the limit is attained is (for a given value of $c$) particularly large for distributions of the Fr\'echet class. We also note
that because $l_{N,N-1} < 1$ for $c < 0$, the normalized correlations are confined to lie between
0 and 1 in this case. This implies that the divergence of $l_{N,N-1}(c)$ for $N \to \infty$ and
$c < 0$, which was demonstrated in Section \ref{Sec:LargeN} to occur for a specific example   
in the Weibull class, cannot happen here. We therefore conjecture that, for distributions in the
Fr\'echet class, the limit (\ref{last}) exists also for $c < 0$.

\subsection{Unified picture}
\label{Sec:Unified}

The results of the preceding subsections can be summarized in a simple scaling picture.
We first recall from \cite{Franke2010} and \cite{LeDoussal2009} that the LDM (with $c>0$)
contains a characteristic time scale $N^\ast$ at which the record rate $p_N(c)$ crosses over from
the i.i.d. behavior  $p_N \approx \frac{1}{N}$ to the limiting value 
$p(c) = \lim_{N \to \infty} p_N(c) > 0$. For the different distributions discussed
above, this time scale diverges for $c \to 0$ according to 
\begin{eqnarray}
N^\ast \propto & c^{-\frac{\xi}{1+\xi}} & \;\;\;\textrm{Weibull class } \label{Nast1} \\
N^\ast \propto & c^{-1} \vert \ln c \vert^{\frac{1}{\beta}-1} & \;\;\;\textrm{Gumbel class } \\
N^\ast \propto & c^{-\frac{\mu}{\mu-1}} & \;\;\;\textrm{ Fr\'echet class } \left(\mu>1\right).
\end{eqnarray}
Ignoring for the moment the logarithmic factor in the Gumbel case, these behaviors can 
be further simplified by expressing the different universality classes in terms of the 
generalized Pareto distribution \cite{Pickands1975}
\begin{equation}
\label{genPareto}
f(x) = \left(1 + \kappa x \right)^{-\frac{\kappa + 1}{\kappa}}.
\end{equation}
For $\kappa > 0$ this is of Pareto type with $\mu = \frac{1}{\kappa}$, for 
$\kappa < 0$ it reduces to a Weibull-type distribution similar to (\ref{Kuramaswamy}) with 
$\xi = - \frac{1}{\kappa}$, and the Gumbel class is represented by the exponential
distribution which arises from (\ref{genPareto}) for $\kappa \to 0$. 
Using this representation, the different cases in (45), (46) and (47) reduce to 
\begin{equation}
\label{Nast2}
N^\ast \propto c^{-\nu} \;\; \textrm{with} \;\; \nu = \frac{1}{1-\kappa}.
\end{equation}
Note that $\nu < 1$ in the Weibull class but $\nu > 1$ in the Fr\'echet class, which explains the slow
convergence to the $N \to \infty$ limit in the latter case.
Moreover, it was shown in \cite{Franke2010} that the integral (\ref{I(N)}) behaves for large $N$ as 
(see also \cite{Sabhapandit2007})
\begin{equation}
\label{I(N)scal}
I(N) \sim  N^{\frac{1}{\nu}-3} = N^{-(2 + \kappa)}.
\end{equation}
To relate this to the behavior of the correlations, we note that for large $N$ Eq.(\ref{taylor_mid})
can be approximately written as 
\begin{eqnarray}
\label{J_asym}
  J(N) & \approx &  - \frac{1}{2} N^4 \frac{\dd}{\dd N}I(N) - N^3 I(N) + {\mathcal{O}(N^2 I(N))}.
\end{eqnarray}
Inserting (\ref{I(N)scal}) we thus conclude that, to leading order,
\begin{equation}
\label{J(N)scal}
J(N) \approx \frac{\kappa}{2} N^3 I(N) \sim N^{1-\kappa},
\end{equation}
which correctly reproduces both the sign and the order of magnitude of the correlations derived
in Section \ref{examples} for the Weibull and Fr\'echet classes: Correlations are positive (negative) for $\kappa > 0$ 
($\kappa < 0$), and they scale sublinearly (superlinearly) with $N$ in the two cases. A similar calculation using the 
refined estimate (\ref{IGumbel}) of $I(N)$ for the Gumbel class shows that the correlations are 
negative (positive) for $\beta > 1$ ($\beta < 1$), in agreement with (\ref{JGumbel}).

\section{Conclusion}
\label{conclusions}

In this paper 
we analyzed the effect of a linear drift on the correlations between records drawn from series of independent RV's, as quantified by 
the normalized joint probability 
$l_{N,N-1}\left(c\right)$. In Section 2 we derived general expressions (exact and approximate) for $l_{N,N-1}\left(c\right)$ and 
recalled the fact that records are independent both for i.i.d. RV's ($c=0$) and in the special case where the 
RV's are drawn from the Gumbel distribution.

Our main analytic results were obtained in Section 3 by way of an expansion in the small $c$ limit, similar to the approach developed previously in \cite{Franke2010}.
Using this approach we were able to show that the correlations are generally negative (`repulsive') for distributions in 
the Weibull class, and positive ('attractive') for distributions in the Fr\'echet class. In the Gumbel class the sign of the 
correlations depends on the stretching exponent $\beta$ in (\ref{fbeta}), with the border between positive and negative correlations
being given by the exponential case $\beta = 1$. In contrast to all other cases, for distributions with an exponential tail 
the correlations are weak and independent of $N$, which is consistent with the fact that this class of distributions also contains
the Gumbel distribution, which has no correlations at all. Simulation results illustrating the different cases are summarized in Fig.~\ref{Fig:comp}.
A special role of the exponential distribution in separating two regimes of qualitatively different behaviors has been noted previously in the
related context of near-extreme events \cite{Sabhapandit2007}. 

Perhaps the most surprising and counterintuitive outcome of our work 
is the discovery of strong positive record correlations for distributions with a power law tail\footnote{We note in this context that in a previous study of records 
drawn from series of independent RV's with an increasing variance only negative correlations were found \cite{Krug2007}.}. 
In view of the substantial interest in detecting and explaining heavy-tailed distributions in all areas of science 
\cite{Sornette,Christensen}, our finding suggests that drift-induced record correlations could be used as a distribution-free
test for detecting power laws or streched exponentials in empirical data \cite{Franke2011}.

An interesting open question concerns the structure of the record correlations in the asymptotic limit $N \to \infty$,
where the record rate approaches a nonzero constant and the record process thus becomes stationary. Based on the work of 
Ballerini and Resnick \cite{Ballerini1985}, we have argued in Section \ref{Sec:LargeN} that the limit (\ref{last}) 
exists for $c>0$ whenever the 
underlying distribution has a finite mean, whereas for $c < 0$ it is possible that $l_{N,N-1}(c)$ diverges for $N \to \infty$. 
The fact that the coefficient of the leading order term in the small $c$ expansion 
generally diverges with $N$ indicates that the limiting function $l^\ast(c)$ may be singular for $c \to 0$, and we have presented
numerical evidence for the occurrence of a discontinuity at $c=0$ for the Weibull class. For heavy-tailed distributions 
in the Fr\'echet class, the large-$N$ asymptotics is difficult to ascertain numerically because of the slow convergence, but we
have argued that in this case $l^\ast(c)$ may exist also for $c<0$. Rigorous work addressing these questions along the lines
of \cite{Ballerini1985} would be most welcome.

\begin{acknowledgements}
This work has been supported by DFG within the \textit{Bonn Cologne Graduate School of Physics and Astronomy}, by Friedrich Ebert Stiftung through a fellowship to GW, and 
by Studienstiftung des Deutschen
Volkes through a fellowship to JF.
\end{acknowledgements}



\end{document}